# A new ground-state band level energy formula for transitional nuclei

Shuifa Shen[1, 2, 3, *], Jing Pan[3], Yupeng Yan[4, 5], Jiejie Shen[6], Feipeng Wang[3], Tingtai Wang[7], Jiaming Jiang[8], Jing Song[3], Mengyun Cheng[3], Lijuan Hao[3]

1 School of Electronic, Electrical Engineering and Physics, Fujian University of Technology, Fuzhou, Fujian 350118, People's Republic of China

2 School of intelligent manufacturing, Zhejiang Guangsha Vocational and Technical University of Construction, Zhejiang, Jinhua 322100, People's Republic of China

3 Hefei Institutes of Physical Science, Chinese Academy of Sciences, Hefei, Anhui 230031, People's Republic of China

4 School of Physics, Institute of Science, Suranaree University of Technology, Nakhon Ratchasima 30000, Thailand

5 Thailand Center of Excellence in Physics (ThEP), Commission on Higher Education, 328 Si Ayutthaya Road, Ratchathewi, Bangkok 10400, Thailand

6 Division of Health Sciences, Hangzhou Normal University, Hangzhou, Zhejiang 310012, People's Republic of China

7 College of science, Zhongyuan University of Technology, Zhengzhou, Henan 450007, People's Republic of China

8 College of Nuclear Science and Engineering, East China University of Technology, Nanchang, Jiangxi 330013, People's Republic of China

**Abstract**: A new three parameter formula is proposed for ground-state bands in even-even soft rotors or called transitional nuclei. The new formula blends those of very soft nuclei and well deformed nuclei. Especially, it is found in fact that the formula proposed by P. von Brentano et al. can be derived from our formula if some approximations are adopted and it can be explained why in their work the energy-dependent term $\beta E$ can be dropped. Compared to their formula it can be resulted in that the range of validity of our formula will be much broader than theirs.

*e-mail: shuifa.shen@inest.cas.cn

Variable-moment-of-inertia (VMI) model

## 1. Introduction

Over the last five decades, the interpretation of energies of ground-state bands in even-even nuclei has been a subject of continuing interest. A variety of theoretical and phenomenological approaches has been developed, ranging from the simplest $I(I+1)$ dependence of the ideal rotor and the $I$ dependence of the ideal harmonic vibrator to numerous more sophisticated expressions designed to account for deviations from these paradigms[1]. In the work of Gupta[2], a perspective of ground-state band energy formulae is given.

In reality, most nuclei lie in between these two limiting symmetries (deformed rotor and spherical vibrator) and are called transitional nuclei, and deviate from either of the above two limiting expressions. In Ref. [3], a three parameter formula for yrast energies in soft rotors or called transitional nuclei was proposed, that is $E = \frac{1}{\Im_0(1+\alpha I+\beta E)} I(I+1)$, where the moment of inertia depends linearly on spin $I$ and excitation energy $E$. However, it was not pointed out how the formula was deduced, which leads one to wonder whether the formula has a clear physical significance. In the present work we develop a new formula for soft rotors or called transitional nuclei.

## 2. Derivation of a new formula for transitional nuclei

On one hand, in 1969 the variable-moment-of-inertia (VMI) model was proposed by Mariscotti et al.[4]. The energy fits in the VMI model were found good for almost all nuclei by them[4], with $R_{4/2}$>2.2. In this model the level energy of the ground-state band is given by the following expression

$$E_I(\Im_I) = \frac{1}{2}[I(I+1)/\Im_I] + \frac{1}{2}C(\Im_I - \Im_0)^2, \quad (1)$$

and the moment of inertia $\Im_I$ (given in unit of $\hbar^2$) for each excited state with spin $I$ is such determined that the level energy of every spin $I$ takes a minimum value, i.e.,

$$\partial E(\Im_I)/\partial \Im_I = 0 \qquad (2)$$

This is called the equilibrium condition in Ref. [4] (My understanding is that the real level energy takes a minimum for the moment of inertia $\Im_I$ because the lower the energy, the more stable it is). $\Im_0$ and C in the above Eq. (1) are the "moment of inertia of the ground-state" and the "restoring force constant", respectively.

From Eq. (1) and Eq. (2) the following relation can be derived

$$\Im_I = \Im_0 / \{1 - [I(I+1)/2C\Im_I^3]\}, \qquad (3)$$

which becomes a cubic equation after a simple transformation

$$\Im_I^3 - \Im_I^2 \Im_0 - [I(I+1)/2C] = 0 . \qquad (4)$$

This cubic equation can be solved analytically, and it has one real root for any finite positive value of $\Im_0$ and $C$. The real root of Eq. (4) is derived as

$$\Im_I = [\frac{1}{3} + \frac{2^{1/3}}{3\left(2 + 27\frac{I(I+1)}{2C\Im_0^3} + 3\sqrt{3}\sqrt{4\frac{I(I+1)}{2C\Im_0^3} + 27[\frac{I(I+1)}{2C\Im_0^3}]^2}\right)^{1/3}} + \frac{\left(2 + 27\frac{I(I+1)}{2C\Im_0^3} + 3\sqrt{3}\sqrt{4\frac{I(I+1)}{2C\Im_0^3} + 27[\frac{I(I+1)}{2C\Im_0^3}]^2}\right)^{1/3}}{3/2^{1/3}}]\Im_0 .$$

Because from Eq. (3) it can be derived that $\Im_I - \Im_0 = \Im_I - \{1 - [I(I+1)/2C\Im_I^3]\}\Im_I = I(I+1)/2C\Im_I^2$, so the following equation for the energy of the state with spin $I$ can be gotten when Eq. (3) is substituted into Eq. (1):

$$E_I = [I(I+1)/2\Im_I]\{1 + [I(I+1)/4C\Im_I^3]\}. \qquad (5)$$

If Eq. (4) is divided by $\Im_0^3$, by defining $r_I = \Im_I / \Im_0$, the following expression is obtained

$$r_I^3 - r_I^2 = \sigma I(I+1), \qquad (6)$$

where $\sigma = \frac{1}{2C\Im_0^3}$. For rigid-body nuclei (the adiabatic limit) $C \to \infty$, so $\sigma = 0$, and hence $r_I = 1$. On the other hand, when the nuclei are at the very soft limit, $\sigma \to \infty$, the left side of Eq. (6) will be $r_I^3 - r_I^2 = r_I^2(r_I - 1) \approx r_I^2 \cdot r_I = r_I^3$, hence Eq. (6) becomes $r_I = [\sigma I(I+1)]^{1/3}$, i.e.,

$$\Im_I = [\sigma I(I+1)]^{1/3}\Im_0 . \quad (7)$$

So in Eq. (5) the term $I(I+1)/4C\Im_I^3 = I(I+1)/4C\sigma I(I+1)\Im_0^3 = 1/4C\sigma\Im_0^3 = \frac{1}{2}$. Hence Eq. (5) then becomes

$$E_I(\sigma \to \infty) = \frac{3}{4}[I(I+1)/\Im_I] = \frac{1}{\frac{4}{3}\sigma^{1/3}[I(I+1)]^{1/3}\Im_0} I(I+1) . \quad (8)$$

On the other hand, in the study of the rotation-vibration coupling energy spectrum in the rotation-vibration model (RVM), an approximate expression for the effective moment of inertia has ever been obtained as follows:

$$\Im_{eff} = \Im_0[1 + bI(I+1)] . \quad (9)$$

For a well-deformed even-even nucleus ($R_{4/2}$=E($4^+$)/E($2^+$)~10/3, i.e., ideal rotor) the level energies of the ground-state band are given by the Bohr-Mottelson (BM) rotation formula

$$E_I = \frac{\hbar^2}{2\Im_0} I(I+1) , \quad (10)$$

therefore $\Im_{eff}$ increases approximately linearly with $E_I$ as follows[5]:

$$\Im_{eff} = C_1 + C_2 E_I . \quad (11)$$

Experimental data on many well-deformed nuclei (150<A<190 and A>220) indicate that this linear relationship between the effective moment of inertia and the excitation energy is fairly well established. In fact, its validity is limited to only well-deformed nuclei ($R_{4/2}$>3.0). For transitional nuclei between spherical and deformed limits neither the vibrator nor rotor limit is very apt. In the present work, a new formula is thus proposed to fit the level energies of ground-state bands in this kind of nuclei. The basis of this expression is simple: it is the ideal rotor expression

$$E = \frac{1}{\Im(I,E)} I(I+1), \qquad (12)$$

but the moment of inertia depends linearly on expression $(I(I+1))^{1/3}$ of spin $I$ and excitation energy $E$. That is

$$\Im(I, E) = \Im_0(1 + \alpha(I(I+1))^{1/3} + \beta E), \qquad (13)$$

where $\alpha$ and $\beta$ are adjustable parameters and $\Im_0$ sets the overall scale. This moment of inertia is the linear superposition of Eq. (8) and Eq. (11), i.e.,

$$a \cdot (\frac{4}{3}\sigma^{1/3}\Im_0[I(I+1)]^{1/3}) + b \cdot (C_1 + C_2 E)$$

$$= bC_1(1 + \frac{a\frac{4}{3}\sigma^{1/3}\Im_0}{bC_1}[I(I+1)]^{1/3} + \frac{C_2}{C_1}E)$$

$$= \Im_0(1 + \alpha(I(I+1))^{1/3} + \beta E), \qquad (14)$$

where $\Im_0$, $a$, and $b$ are three parameters. The idea of our present work is just like that of Ref. [6], where the empirically based expression $E(I) = aI + bI(I+1)$ is applied to blend vibrator and rotor concepts. The parameters in Eq. (13) can be fixed by fitting to yrast (ground-state) band below the band crossing data for the $2^+$, $4^+$ and $6^+$ energies, where Eq. (13) predicts all the higher spin levels, or by doing a least squares fit to the entire (pre-backbending or alignment portion of the) band or quasiband.

In Ref. [3], a three parameter formula for yrast energies in soft rotors or called transitional nuclei was also proposed, that is, $E = \frac{1}{\Im_0(1 + \alpha I + \beta E)} I(I+1)$. In our formula in Eq. (13) the term $\alpha(I(I+1))^{1/3} \approx \alpha I^{2/3}$ at high spin $I$, but it still differs largely from $\alpha I$. In addition, the energy-dependent term $\beta E$ was dropped in Ref. [3] since their test fits show that it is very small in the transitional region. However, in our formula it may not be the case. In fact, the terms $\alpha(I(I+1))^{1/3}$ and $\beta E$ should be in the same status. Through careful consideration, it is found that their formula can be deduced from our formula as follows if some approximations are adopted:

$$\begin{aligned}\Im(I,E) &= \Im_0(1+\alpha(I(I+1))^{1/3}+\beta E) \\ &\approx \Im_0(1+\alpha(I(I+1))^{1/3}+\beta' I(I+1)) \\ &\approx \Im_0(1+\alpha I^{2/3}+\beta' I^2) \\ &\approx \Im_0(1+\alpha' I), \qquad (15)\end{aligned}$$

where $\beta E \approx \beta' I(I+1)$ because for ideal rotor Eq. (10) can be used and $\alpha I^{2/3} + \beta' I^2 \approx \alpha' I$, this is a very rough approximation. Here in fact $\Im_0(1+\alpha' I)$ is identical to $\Im_0(1+\alpha I)$ because $\alpha'$ and $\alpha$ are both only the parameters, so this can explain why in Ref. [3] the energy-dependent term $\beta E$ could be dropped.

Through a literature survey, it is found that the Eq. (3) of Ref. [3] is exactly the Eq. (6) of Ref. [7], namely, $E = AJ(J+1)/(1+\sigma_1 J)$. So the moment of inertia of formula $E = \frac{1}{\Im_0(1+\alpha I+\beta E)} I(I+1)$ can be regarded as the linear superposition of Eq. (6) of Ref. [7] and Eq. (11), i.e.,

$$\begin{aligned}& a\cdot(\frac{1}{A}(1+\sigma_1 J)) + b\cdot(C_1 + C_2 E) \\ &= (\frac{a}{A} + bC_1) + \frac{a\sigma_1}{A} J + bC_2 E \\ &= \Im_0(1+\alpha I+\beta E). \qquad (16)\end{aligned}$$

In Ref. [7] it is pointed out by Gupta that for the softer (i.e. both the spherical and neutron-deficient) nuclei the agreement between the experimental data and the ones predicted by the equation $E = AJ(J+1)/(1+\sigma_1 J)$ is not very satisfactory since the first-order nuclear softness $\sigma_1$ is shown to be required even for the strongly deformed nuclei, and Eq. (11) is only applicable for the well-deformed nuclei as already mentioned above. So it can result in that Eq. (2) of Ref. [3] can't be applicable for transitional region 1 ($2.4 \le R_{4/2} \le 2.7$)[8] although it is said that this formula is particularly successful in soft rotors with $2.8 \le R_{4/2} \le 3.2$[3] whereas our formula can be applicable for broader region because the moment of inertia of our formula is the linear superposition of those of very soft nuclei ($R_{4/2} \sim 2.23$)[4] and well-deformed ones.

## 3. Summary

In summary, in the present work, a new formula is proposed to fit the level energies of ground-state bands in even-even transitional nuclei and the physical significance of the formula proposed by P. von Brentano et al. is given. The derivation is detailed and the underlying assumptions are reasonably well argued. Our formula can be applicable for broader region than the one proposed by P. von Brentano et al., especially it is found that their formula can be deduced from our formula if some approximations are adopted and it is explained why in their formula the energy-dependent term $\beta E$ could be dropped.

References

[1]N. V. Zamfir, R. F. Casten, Phys. Rev. Lett. 75, 1280 (1995)

[2]J. B. Gupta, Pramana–J. Phys. 89, 34(2017)

[3]P. von Brentano, N. V. Zamfir, R. F. Casten et al., Phys. Rev. C69, 044314(2004)

[4]M. A. J. Mariscotti, G. Scharff-Goldhaber, and B. Buck, Phys. Rev. 178, 1864(1969)

[5]P. Holmberg and P. O. Lipas, Nucl. Phys. A117, 552(1968)

[6]H. Ejiri, M. Ishihara, M. Sakai, K. Katori, and T. Inamura, J. Phys. Soc. Jpn. 24, 1189(1968)

[7]R. K. Gupta, Phys. Lett. B36, 173(1971)

[8]D. Bonatsos, A. Klein, Phys. Rev. C29, 1879(1984)